\documentclass[10pt,aps,prl,twocolumn,superscriptaddress,preprintnumbers,floatfix,
notitlepage,nofootinbib]{revtex4-1}

\usepackage{amsmath,amsfonts,amssymb}
\usepackage{fancyhdr}
\usepackage{graphicx}
\usepackage{xspace}
\usepackage{rotating}
\usepackage[normalem]{ulem}
\usepackage{braket}
\usepackage{verbatim}
\usepackage{xcolor}
\usepackage{hyperref}
\usepackage[utf8]{inputenc}
\def\apjl{The Astrophysical Journal Letters}

 \usepackage{siunitx}
\usepackage{pgfplots}

 \usepackage{tabularray}
\usepackage{tikz}
\usetikzlibrary{shapes.geometric, arrows}

\tikzstyle{startstop} = [rectangle, rounded corners, 
minimum width=1cm, 
minimum height=1cm,
text centered, 
draw=black, 
fill=red!30]

\tikzstyle{io} = [circle,  
radius=2cm, 
text centered, 
draw=black, fill=blue!30]

\tikzstyle{io0} = [circle,  
radius=2cm, 
text centered, 
draw=black, fill=orange!30]

\tikzstyle{prd0} = [rectangle, 
minimum width=2m, 
minimum height=2cm, 
text centered, 
text width=2cm, 
draw=black, 
fill=orange!30]

\tikzstyle{pbh} = [diamond, 
minimum width=2cm, 
minimum height=1cm, 
text centered, 
draw=black, 
fill=green!30]
\tikzstyle{arrow} = [thick,->,>=stealth]

\usepackage{calligra}
\DeclareMathAlphabet{\mathcalligra}{T1}{calligra}{m}{n}
\DeclareFontShape{T1}{calligra}{m}{n}{<->s*[2.2]callig15}{}

\usepackage[mathscr]{euscript}
\usepackage[export]{adjustbox}

\def\l{\left}
\def\r{\right}

\def\beq{\begin{equation}}
\def\eeq{\end{equation}} 
\def\be{\begin{eqnarray}}
\def\ee{\end{eqnarray}}

\def\MPBH{M_{\rm PBH}}

\definecolor{lime}{HTML}{A6CE39}
\DeclareRobustCommand{\orcidicon}{
	\begin{tikzpicture}
	\draw[lime, fill=lime] (0,0) 
	circle [radius=0.2] 
	node[white] {{\fontfamily{qag}\selectfont \tiny ID}};
	\draw[white, fill=white] (-0.0625,0.095) 
	circle [radius=0.007];
	\end{tikzpicture}
	\hspace{-2mm}
}

\foreach \x in {A, ..., Z}{\expandafter\xdef\csname orcid\x\endcsname{\noexpand\href{https://orcid.org/\csname orcidauthor\x\endcsname}
			{\noexpand\orcidicon}}
}

\begin{document}

\title{Ultra-low mass PBHs in the early universe can explain the PTA signal}
\author{Nilanjandev Bhaumik\orcidA{}}
\email{nilanjandevbhaumik@gmail.com}
\author{Rajeev Kumar Jain\orcidB{}}%
 \email{rkjain@iisc.ac.in}
\affiliation{%
Department of Physics, Indian Institute of Science, \\
C. V. Raman Road, Bangalore 560012, India
}

\author{Marek Lewicki\orcidC{}}
 \email{marek.lewicki@fuw.edu.pl}
\affiliation{
Institute of Theoretical Physics, Faculty of Physics, University of Warsaw, ul. Pasteura 5, 02-093 Warsaw, Poland
}%

\date{\today}

\begin{abstract}
Pulsar Timing Array collaborations have recently announced the discovery of a stochastic gravitational wave background (SGWB) at nanohertz frequencies. We analyze the GW signals from the domination of ultra-low mass primordial black holes (PBHs) in the early universe and show that they can explain this recent discovery.
This scenario requires a relatively broad peak in the power spectrum of scalar perturbations from inflation with a spectral index in a narrow range of $1.45$ to $1.6$. The resulting PBH population would have mass around $10^{8}$g, 
and the initial abundance $\beta_f$ lies between $10^{-10}$ and $10^{-9}$. We find that this explanation is preferred by the data over the generic model, assuming supermassive BHs as the source. 
These very light PBHs would decay before Big Bang Nucleosynthesis (BBN); however, upcoming third-generation terrestrial laser interferometers would be able to test the model by observing the GW spectrum produced during the formation of the PBHs. Also, the scalar power spectra associated with our scenario will be within the reach of PIXIE probing CMB spectral distortions.

\end{abstract}

\maketitle


{\bf Introduction:}
In a recent data release, multiple Pulsar Timing Array (PTA) experiments have reported evidence for a stochastic gravitational wave background (SGWB) including NANOGrav~\cite{NANOGrav:2023gor, NANOGrav:2023hde}, EPTA (including the data from InPTA)~\cite{Antoniadis:2023lym, EPTA:2023fyk}, PPTA~\cite{Zic:2023gta, Reardon:2023gzh}, and CPTA~\cite{Xu:2023wog}. The most obvious explanation for such a background would be from supermassive black hole binary mergers.
However, at this stage, it is impossible to determine whether the origin is astrophysical~\cite{NANOGrav:2023hfp, Antoniadis:2023aac, Ellis:2023dgf} or if this is one of the possible signals from the early universe~\cite{NANOGrav:2023hvm, Antoniadis:2023zhi}. Cosmological SGWB sources discussed in the literature to explain NANOGrav include the SGWB from the massive primordial black hole (PBH) formation models~\cite{ Inomata:2023zup, Franciolini:2023pbf, Cheung:2023ihl, Balaji:2023ehk, Firouzjahi:2023lzg, Unal:2023srk, Frosina:2023nxu, Liu:2023ymk, HosseiniMansoori:2023mqh, Liu:2023pau}, PBH mergers~\cite{Depta:2023qst, Gouttenoire:2023nzr}, cosmological phase transition~\cite{Salvio:2023ynn, Gouttenoire:2023bqy, Ghosh:2023aum, An:2023jxf, Jiang:2023qbm, Athron:2023mer, DiBari:2023upq, Li:2023bxy}, cosmic strings and domain walls~\cite{Ellis:2023tsl, Lazarides:2023ksx, Zhang:2023nrs, Yamada:2023thl, Lu:2023mcz, Babichev:2023pbf, Ge:2023rce, Li:2023tdx, Kitajima:2023cek, King:2023cgv, Lazarides:2023rqf}, models of axion inflation~\cite{Unal:2023srk, Niu:2023bsr, Murai:2023gkv}, blue tilted inflationary tensor spectrum~\cite{Vagnozzi:2023lwo, Borah:2023sbc, Datta:2023vbs, Choudhury:2023kam} and etc.

In this paper, we study a scenario wherein a peaked spectrum of inflationary first-order scalar perturbations leads to the formation of ultra-low mass PBHs in the post-inflationary radiation-dominated universe. These PBHs overcome the radiation energy density after some time and dominate the expansion history of the universe until they evaporate due to Hawking radiation. As PBHs in such a scenario evaporate and contribute to subsequent radiation domination (RD) far before the Big Bang Nucleosynthesis (BBN), the later dynamics leading to cosmic microwave background (CMB) are not altered. Yet, a PBH-dominated era can lead to a resonant amplification in the SGWB~\cite{Inomata:2019ivs}. Earlier works studied such mechanisms of SGWB generation both for inflationary adiabatic scalar power spectrum~\cite{Inomata:2020lmk} and isocurvature-induced adiabatic scalar perturbations contributed by PBH density fluctuations~\cite{Papanikolaou:2020qtd, Domenech:2020ssp, Papanikolaou:2022chm} ( also see \cite{Basilakos:2023xof} explaining NANOGrav signal in this context ), which leads to a doubly peaked SGWB spectrum~\cite{Bhaumik:2022pil, Bhaumik:2022zdd}. Here we consider an integrated picture with a peaked inflationary scalar power spectrum appropriate for PBH formation in the corresponding mass range and abundance and find that the resulting resonant SGWB has the potential to explain the recent PTA signals.

{\bf PBH formation from amplified inflationary scalar power spectrum:}
Many inflationary models have been proposed to produce an appropriate amplification in the small-scale scalar perturbations, leading to the formation of PBHs in the post-inflationary era. These amplified perturbation modes collapse when they re-enter the horizon during the  RD~\cite{Germani:2017bcs, Garcia-Bellido:2017mdw, Bhaumik:2019tvl} or the reheating era \cite{Bhattacharya:2019bvk}. 

Here we will focus on the formation of ultra-low mass PBHs, which would have evaporated before BBN due to Hawking radiation~\cite{Hawking:1974rv}. We assume the formation of these PBHs  ($M_{\rm PBH} < 10^9 {\rm g}$) during the early radiation domination (eRD) era from an amplified inflationary power spectrum, subsequent PBH domination or early matter domination (eMD) and their evaporation leading to standard RD. Such a scenario requires a peak in the inflationary power spectrum at a very small scale. In single field inflation for potentials with a local extremum \cite{Germani:2017bcs, Garcia-Bellido:2017mdw, Bhaumik:2019tvl} or for varying sound speed \cite{Kamenshchik:2018sig, Cai:2018tuh, Zhai:2022mpi}, it is possible to generate a large enhancement of the scalar perturbations. We model this peak in the inflationary scalar power spectrum as a broken power law,
\begin{align}\label{eq:PR}
{\mathcal P}_{\mathcal R}=A_s \left(\frac{k}{k_{\rm p}}\right)^{n_s-1} +{A}_{0}\left\{
\begin{aligned}
&\left(\frac{k}{k_{\rm pk}}\right)^{n_0-1}\hskip0.1em &k\leq k_{\rm pk}\\
&\left(\frac{k}{k_{\rm pk}}\right)^{-2}\hskip0.1em & k\geq k_{\rm pk}
\end{aligned}
\right.
\end{align}
where we take the scalar spectrum amplitude $A_s =2.1 \times 10^{-9}$, scalar index $n_s=0.965$, and pivot scale $k_p=0.05\, { \rm Mpc^{-1}}$ from Planck-2018 data~\cite{Planck:2018jri}. The value of $A_0$ and $k_{\rm pk}$ determines the height and location of the peak and, thus, the abundance and mass range of produced PBHs, while $n_0$ reflects the slope of the blue-tilted part of the power spectrum, with a theoretical upper bound, $n_0 \lesssim 5$ \cite{Byrnes:2018txb, Ozsoy:2019lyy, Carrilho:2019oqg}.

For PBHs forming in RD, their mass can be related to the Horizon mass of that time as,
\be
\MPBH(k) \equiv \gamma M_H = \frac{4 \pi}{3} \frac{\gamma \rho}{H^{3}}\biggm \vert_{k=aH} \, , 
\ee
where we took into account the appropriate efficiency factor $\gamma \approx 0.2$ \cite{Harada:2013epa}.
The initial PBH mass fraction can be estimated as,
\begin{equation}
\beta_{f}(\MPBH)
=\frac{1}{2}{\rm erfc}\l(\frac{\delta_c}{\sqrt{2}\, \sigma_\delta (\MPBH(R))}\r).     \label{ps01}
\end{equation}
Here we use the simple Press-Schechter formalism for estimating PBH initial mass fraction $\beta_f$ 
with critical density contrast $\delta_c$, and the variance of the density contrast $ \sigma_\delta^2 $ coarse-grained at a comoving scale $R$~\cite{Bhaumik:2019tvl}.
It is important to note that $\beta_f$ is exponentially sensitive to the peak height of the scalar power spectrum ($A_0$), which makes it necessary that one chooses a fine-tuned value of $A_0$ to avoid negligible or too large production of PBHs. This fine-tuning problem is generic when considering the PBH formation in RD~\cite{Nakama:2018utx}. 

The estimation of $\beta_f$ involves many uncertainties as discussed in Appendix B of \cite{Bhaumik:2019tvl}. One of them concerns the value of the critical density contrast $\delta_c$. This threshold, in principle, should depend on the shape of the assumed power spectrum of curvature perturbations~\cite{Musco:2018rwt, Musco:2020jjb}; however, for simplicity, we use a constant analytical value for $\delta _c$ derived in
\cite{Harada:2013epa}. Moreover, non-Gaussian effects can also prove to be quite important in the computation of the PBH abundance. While the PBH forming models typically predict a highly non-Gaussian tail for the curvature perturbation distribution, another source of non-Gaussianity is the non-linear relationship between density contrast and curvature perturbation~\cite{Young:2019yug, DeLuca:2019qsy}. As we do not consider any particular inflationary model here, the more accurate estimation of the combined effects of non-Gaussianities from both these origins is outside the scope of this paper, and we leave it for future work.

The PBHs act as non-relativistic matter whose relative energy density grows proportional to the scale factor during eRD. Thus, ultra-low mass PBHs produced during eRD at conformal time $\tau=\tau_f$ can dominate the universe
at $\tau=\tau_m$ before PBH evaporation at $\tau=\tau_r$.
The comoving horizon sizes at those different transition points can be expressed in terms of PBH mass $\MPBH$ and the initial abundance $\beta_f$~\cite{Bhaumik:2022pil},
\begin{eqnarray}
k_{r} &=& \frac{1}{\tau_r} \approx 2.1 \times 10^{11} \left( \frac{M_{\rm PBH}}{10^4 {\rm g} } \right)^{-3/2} \text{Mpc}^{-1} , \\
 k_{m} &=& \frac{1}{\tau_m} \approx 3.4 \times 10^{17} \left( \frac{M_{\rm PBH}}{10^4 {\rm g} } \right)^{-5/6} \beta_f^{2/3}\, \text{Mpc}^{-1},  \\
 k_{f}&=&\frac{1}{\tau_f} \approx3.4 \times 10^{17} \left( \frac{M_{\rm PBH}}{10^4 {\rm g} } \right)^{-5/6} \beta_f^{-1/3}\, \text{Mpc}^{-1} .
 \label{waves}
\end{eqnarray}
We quantify the duration of PBH domination with the ratio of conformal times at the end and the start of PBH domination, $\tau_{\rm rat} \equiv \tau_r /\tau_m$.

{\bf Evolution of first-order scalar perturbations and resulting SGWB:}
The SGWB generated at the time of formation of massive PBHs due to the amplified scalar perturbations sourcing the tensor perturbation at second order is the most popular mechanism, discussed extensively in the literature \cite{NANOGrav:2023hvm, Inomata:2023zup, Franciolini:2023pbf}.
In the case of ultra-low mass PBHs, SGWB associated with PBH-formation would peak at a very high frequency which is incompatible with the NANOGrav frequency band ($\approx  10^{-9}\, {\rm Hz}$), but the resonant SGWB, generated during the very onset of RD, due to the nontrivial evolution of first-order scalar perturbations during eMD can lead to an SGWB amplification around NANOGrav frequency.

The existence of two nontrivial phases before standard RD determines the evolution of scalar perturbation modes, which re-enter the horizon during eRD and eMD. While the amplitude of first-order scalar perturbation modes, which re-enter the horizon during eRD, are suppressed rapidly \cite{Mukhanov:2005sc}, the modes re-entering the horizon during eMD stay nearly constant. These constant subhorizon modes oscillate rapidly with high frequency and amplitude once the RD era starts, leading to resonant amplification in the SGWB~\cite{Inomata:2019ivs}.

The energy density per logarithmic $k$ interval $\Omega_{\rm GW}(\tau,k)$ for the second-order SGWB sourced by first-order scalar perturbations can be expressed as, 
\be
\label{omega_uv}
\Omega_{\rm GW}(\tau,k)= 
\frac{1}{6}  \int_{0}^{\infty}dv \int_{|1-v|}^{1+v} du
\left(\frac{4 v^2-(1+v^2-u^2)^2}{4 u v}\right)^2 \nonumber\\ 
\times \hspace{0.5cm}\overline {\cal I}^2(v,u,x) {\cal P}_{\cal R}(kv){\mathcal P}_{\mathcal R}(ku) \hspace{1cm}\,  
\ee
For the SGWB from PBH formation, the last two terms are the inflationary scalar perturbation spectra (at $\tau=\tau_i$) and the kernel $\overline{\cal I}^2 \equiv \overline{\cal I}_{\rm eRD}^2 $ refers to the evolution of scalar perturbation modes during eRD, $x= k\tau$ is dimensionless time variable.
On the other hand, for the resonant SGWB generated during RD, the last two terms are the first-order scalar perturbation spectra at the very start of RD and $\overline{\cal I}^2 \equiv \overline{\cal I}_{\rm RD}^2 $ comes from the evolution of first-order scalar perturbation modes during RD \cite{Inomata:2020lmk, Bhaumik:2020dor}.
Using Eqn. (\ref{omega_uv}), we estimate SGWB from PBH formation $\Omega_{\rm GW}^{\rm form}(\tau_m,k)$ at $\tau=\tau_m$  and resonant SGWB, $\Omega_{GW}^{\rm res} (\tau_l,k)$ at $\tau =\tau_l$, which corresponds to a late time during RD, by which time the scalar perturbation source stops contributing to the kernel. During eRD, we can use the pure RD expressions for the kernel \cite{Espinosa:2018eve, Kohri:2018awv},
\be
\!\!\!\!\!\!\!\!\!\!\!\!
\overline {\cal I}_{\rm eRD}^2(v,u,x \to \infty) 
=  \frac{1}{2}\l(\frac{3 (u^2+v^2-3)}{4u^3v^3}\r)^2\nonumber \\
\biggl[\l( -4uv +(u^2+v^2-3)\,{\rm log}\l|\frac{3-(u+v)^2}{3-(u-v)^2}\r|\r)^2 \biggr. \nonumber \\
 \biggl. \!\! +\pi^2 (u^2+v^2-3)^2 \Theta(u+v-\sqrt{3})\biggr].
 \label{ird}
\ee
Due to the presence of an eMD before the late RD phase, the expression of $\overline{\cal I}_{\rm RD}^2 $ is more involved. As we limit ourselves to a finite duration of PBH-dominated era ($\tau_{m} \ll \tau_{r}$), we can take the scale factor during RD $a_{\rm RD} \propto (\tau - \tau_r/2)$~\cite{Bhaumik:2022pil} and use the expression for  $\overline {\cal I}_{\rm RD}^2$  derived in Appendix A of our earlier work~\cite{Bhaumik:2020dor} where the peculiar velocities of PBHs are assumed to have negligible contribution in comparison to other terms. It is important to note that, this expression is more general and can be reduced to pure RD expression in the limit $\tau_r \to 0$, as shown in the Appendix A of ~\cite{Bhaumik:2020dor}.

The next step is to estimate the present day ($\tau=\tau_0$) value of $\Omega_{\rm GW}$ for both cases,
\be
\Omega_{\rm GW}^{\rm form}(\tau_0,k) = \l(\frac{a_m}{a_r}\r) c_g ~\Omega_{r,0} ~\Omega_{\rm GW}^{\rm form}(\tau_m,k)\, ,
\label{oform}
\ee
\be
\Omega_{\rm GW}^{\rm res}(\tau_0,k) =  c_g ~\Omega_{r,0} ~\Omega_{\rm GW}^{\rm res}(\tau_l,k)\, ,
\ee
where $\Omega_{r,0}$ is the present radiation energy density, $c_g \approx 0.4$ if we take the number of relativistic degrees of freedom to be $\sim 106.7$~\cite{Espinosa:2018eve}. 
The extra factor in Eqn. (\ref{oform}); the ratio of scale factors $a_m/a_r$ comes from the dilution of the SGWB energy density during the PBH-dominated era.

When the PBH domination or eMD starts, scalar modes with comoving wavenumber  $k_r < k < k_m $ re-enter the horizon during eMD and stay nearly constant. As a result, at the start of RD, we get the scalar spectra retaining their inflationary shape for co-moving wavenumber $k< k_m$, while $k> k_m$ part of the power spectra gets significantly suppressed, setting a cutoff scale around $k=k_m$. 
Thus, in the case of broad inflationary power spectra where the tail part of the peak continues below $k < k_m$, the resultant scalar power spectra at the start of RD contain a blue-tilted amplified region (Shown in the left panel of Fig. \ref{fig3}), which in turn leads to a higher amplification in the resonant SGWB spectra, comparable with NANOGrav 15 years observation.
To estimate the resonant SGWB, we assume the transition from PBH domination to RD to be nearly instantaneous and follow the formalism developed in \cite{Inomata:2019ivs}. It is interesting to note that, in this resonant amplification scenario, both the duration of the eMD and the amplitude of scalar power spectra at the start of eMD play important roles in determining the resultant SGWB spectra.

\begin{table}
  \begin{center}
    \begin{tblr}{|Q[c,1.6cm]|Q[c,2.5cm]|Q[c,2.7cm]|}
   \hline  
   \bf{Parameter}  &{\bf{Description}} & \SetCell[c=1]{c}{{\bf{Prior}}} \\
\hline
\SetCell[r=1]{c}
$\log_{10}(\frac{\MPBH}{~1~{\rm g}})$ & PBH mass in logscale &  
\SetCell[c=1]{c}{{{Uniform $[7.5, 8.5]$}}} \\\hline
$\tau_{rat}$ & Duration of PBH domination &  \SetCell[c=1]{c}{{Uniform $[100, 300]$}} \\\hline
$n_0$ & Tilt of the scalar spectra &  \SetCell[c=1]{c}{{Uniform $[1.4, 1.6]$}} \\\hline
    \end{tblr}
\caption{\label{tab:priors} Parameter priors used in the Bayesian analysis of this work.}
  \end{center}
\end{table}

\begin{table}
  \begin{center}
    \begin{tblr}{|Q[c,1.7cm]|Q[c,1.7cm]|Q[c,2.0cm]|Q[c,1.6cm]|}
      \hline
      \SetCell[r=2]{c}{{{\textbf{Model} }}}&\SetCell[r=2]{c}{{{\textbf{Parameters} }}}
    & 
\SetCell[c=2]{c}{{{\textbf{Posterior mean}}}} \\
 & 
 \hline
 &
 \textrm{NG15} &
 \textrm{IPTA2}\\
      \hline \hline 
          \SetCell[r=3]{c}{{{Ultra-low mass PBH model}}}& $\log_{10}(\frac{\MPBH}{1 {\rm g}}) $ & $ 7.99^{+0.13}_{-0.15}$ & $8.11^{+0.35}_{-0.12}$   \\\hline 
                   & $\log_{10}(\beta_f)$& $ -9.57^{+0.15}_{-0.11}$ &$-9.69^{+0.22}_{-0.28}$  \\\hline 
                   & $n_0$& $1.503^{+0.025}_{-0.042}$ &$1.507^{+0.047}_{-0.040}$   \\
         \hline 
    \end{tblr}
    \caption{\label{tab:meanPoseterior}  Mean and 68$\%$ confidence interval values from the probability distribution of ultra-low mass PBH model parameters.}
  \end{center}
\end{table}

\begin{table}
  \begin{center}
    \begin{tblr}{|Q[c,1.5cm]|Q[c,1.65cm]|Q[c,1.6cm]|Q[c,1.5cm]|}
      \hline
      \SetCell[r=2]{c}{{{\textbf{Model X} }}}
    & \SetCell[r=2]{c}{{{\textbf{Model Y} }}} &  
        \SetCell[c=2]{c}{{{${\textrm{BF}_{Y,X}}$}}} \\
 &&
 \hline
 \textrm{NG15} &
 \textrm{IPTA2}\\
        \hline
         SMBHB \hspace{0.5cm} & Ultra-low mass PBHs  & $18.00\pm 1.75$ & $3.31\pm 0.09$ \\
      \hline
    \end{tblr}
    \caption{\label{tab:BF} Bayesian factors $\rm BF_{Y,X}$ with values  exceeding 1 shows support for model $Y$ with respect to model $X$. We can see that ultra-low mass PBH model-induced resonant SGWB is favored with respect to SGWB from mergers of SMBH binaries in NG15 and IPTA2 data.}
  \end{center}
\end{table}

\begin{figure}[t]
\label{fig5}
\includegraphics[scale=0.6]{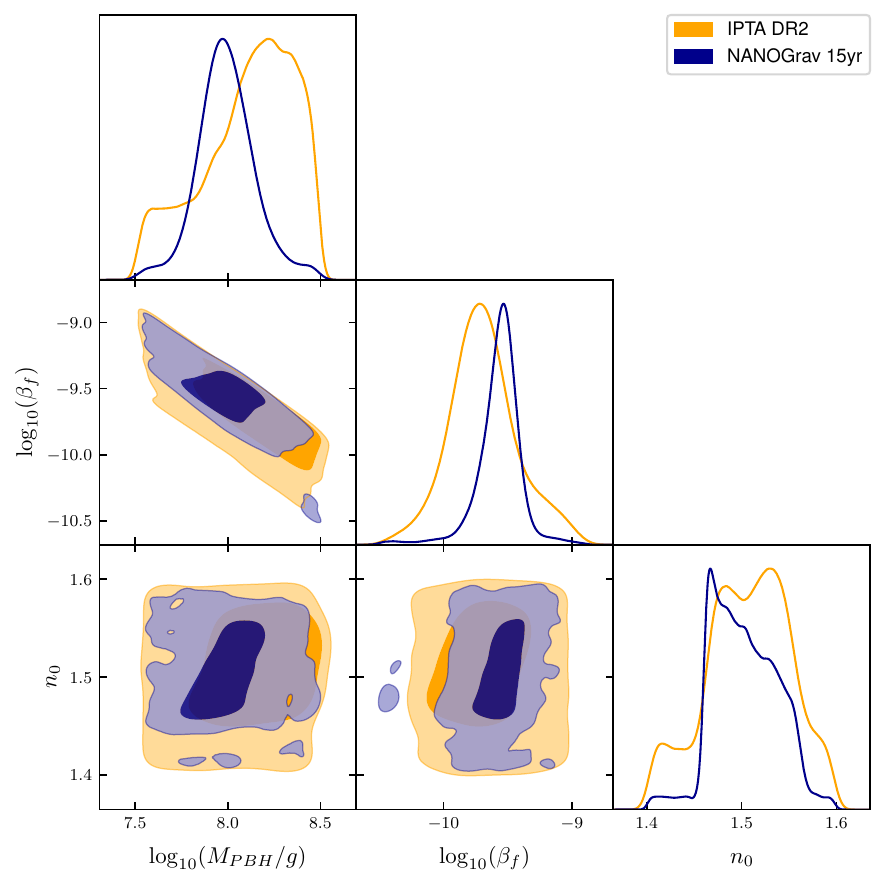} 
\caption{\label{fig:scan} {Triangular plot from the MCMC runs for our model parameters with NANOGrav 15 yr (blue) and IPTA DR2 (orange).}}
\end{figure}

\begin{figure}[h]
\centering
\begin{tikzpicture}[node distance=2cm]
\node (start2) [io] { \hspace{0.1cm} $\tau_{\rm rat}$ \hspace{0.1cm}  };
\node (start) [io,, left of=start2] {$M_{\rm PBH}$};

\node (start3) [io, right of=start2] { \hspace{0.25cm} $n_0$ \hspace{0.25cm}};
\node (in1) [startstop, below of=start2] {$\beta_f$};
\node (in2) [startstop, left of=in1] {$k_f$, $k_m$, $k_r$};
\node (pro1) [startstop, below of=in1] {$A_0$};
\node (pro2a) [io0, right of=pro1, yshift=1cm, xshift=2cm] {$\Omega^{\rm res}_{\rm GW} h^2$};
\node (pro2b) [pbh, right of=pro1, xshift=-4cm] {$\delta_c$};
\draw [arrow] (start) -- (in1);
\draw [arrow] (start2) -- (in1);
\draw [arrow] (start) -- (in2);
\draw [arrow] (in1) -- (in2);
\draw [arrow] (in1) -- (pro1);
\draw [arrow] (start3) -- (pro1);
\draw [arrow] (start3) -- (pro2a);
\draw [arrow] (in2) -- (pro2a);
\draw [arrow] (pro1) -- (pro2a);
\draw [arrow] (pro2b) -- (pro1);
\end{tikzpicture}
\caption{A schematic view of the steps involved in our analysis}
\label{flc}
\end{figure}
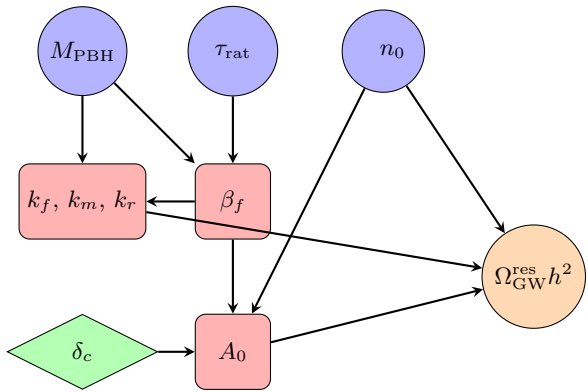

Prolonged duration of eMD can lead to density contrast of $\mathcal{O}(1)$ and in such a scenario, the estimation of SGWB based on linear order scalar perturbations on small scales ($k \gtrsim k_{\rm nonlinear}$) might become inaccurate~\cite{Assadullahi:2009nf, Inomata:2020lmk}. 
This also opens up the possibility of another population of PBHs to form during eMD, however, recent investigations taking into account the velocity dispersion generated during a non-linear era~\cite{Harada:2022xjp} suggest that their abundance would be negligible.

\begin{figure*}
\includegraphics[width=0.5\textwidth]{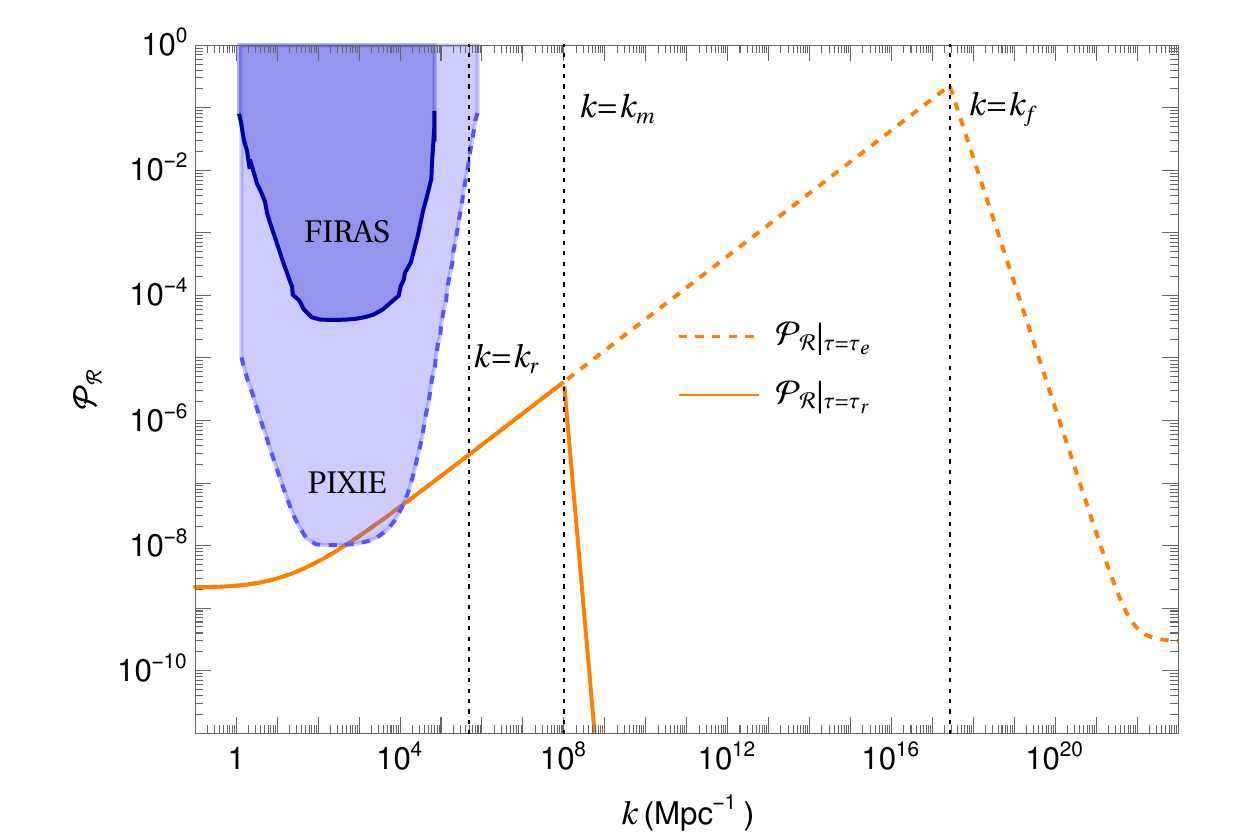} 
\includegraphics[width=0.45\textwidth]{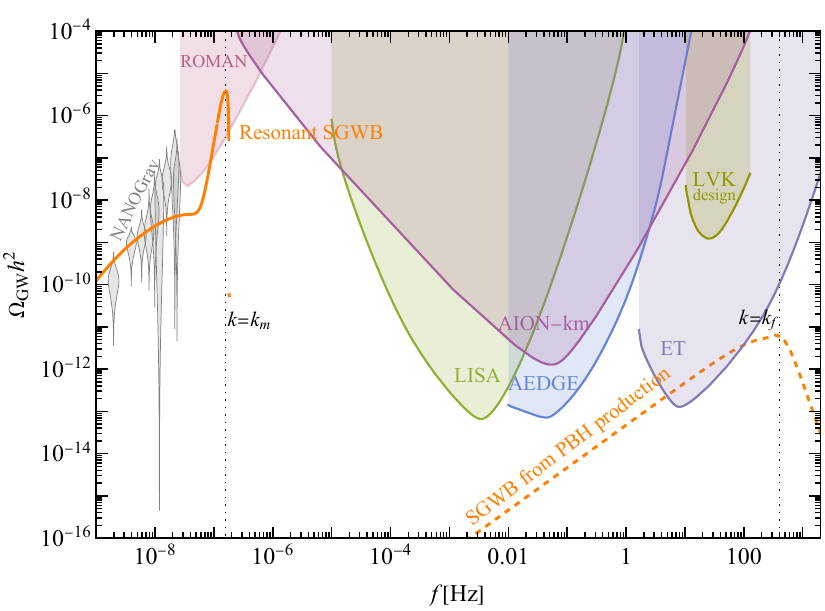} 
\caption{\label{fig3} {\textbf{Left:} The time evolution of first-order scalar power spectrum ${\cal P}_{\cal R}$ at the end of inflation, $\tau=\tau_e$ (Dashed) and at the start of RD $\tau=\tau_r$ (solid) for the posterior mean values from our scan with NANOGrav 15 years data. We also added the CMB spectral distortion bound from FIRAS \cite{1994ApJ...420..439M, 2012ApJ...758...76C, 2014PhRvL.113f1301J} and projected sensitivity for future measurements like PIXIE \cite{ 
2011JCAP...07..025K,2021ExA....51.1515C}.} \hspace{0.2cm} 
\textbf{Right:} { Associated SGWBs: SGWB generated during eRD from PBH formation (dashed) and resonant SGWB generated at the very start of RD due to the nontrivial evolution of scalar perturbations during eMD (solid). 
The vertical lines refer to the comoving horizon sizes at the start of PBH domination $k_m$, the start of RD after PBH evaporation $k_r$, and at the time of PBH formation $k=k_f$.
}}
\end{figure*}

{\bf Signal analysis:}
Next, we will quantify how well our resonant SGWB originating from broadly peaked inflationary scalar perturbations can explain the PTA observations.
We use NG15 dataset~\cite{NANOGrav:2023gor,NG15yrdata} and version B of the IPTA2 dataset~\cite{Antoniadis:2022pcn, IPTADR2data}.  Our Bayesian analysis of both IPTA2 and NG15 data rely on  ${\tt PTArcade}$ \cite{Mitridate:2023oar} in ${\tt enterprise}$~\cite{enterprise} mode without Hellings and Downs (HD) \cite{1983ApJ...265L..39H} correction. Given the observed PTA data $\mathcal{D}$, using Bayes's theorem, we can express the likelihood function in terms of the posterior distribution, $\mathcal{P}(\theta|\mathcal{D})$ for model parameters $\theta$,
 \begin{equation}
     \mathcal{P}(\theta|\mathcal{D}) = \frac{\mathcal{P}(\mathcal{D}|\theta)\mathcal{P}(\theta)}{\mathcal{P}(\mathcal{D})}.
 \end{equation}
Here, $\mathcal{P}(\theta)$ denotes the prior distribution, while $\mathcal{P}(\mathcal{D})$ is the marginal likelihood, used as a normalization constant so that the integration of the posterior distribution is unity.
We estimate the marginal likelihood for PTA data ($\mathcal{D}$), in support of model $Y$, versus model $X$, with the Bayesian factor,
\begin{equation}
    \textrm{BF}_{Y,X} \equiv \mathcal{P}(\mathcal{D}|Y)\,/\,\mathcal{P}(\mathcal{D}|X),
\end{equation}
and use ${\tt GetDist}$~\cite{Lewis:2019xzd} to plot the results.  We confine ourselves to 14 and 13 frequency bins of the NG15 and IPTA2 datasets to avoid pulsar-intrinsic excess noise. 
We simultaneously analyze the supermassive black hole binary (SMBHB) model, with theoretical priors \cite{Mitridate:2023oar} as model X, the reference model for our Bayesian analysis. For modeling the SMBH parameter priors, we have chosen {\texttt{smbhb}}\texttt{=True} option in \texttt{PTArcade}, which adds expected signal produced by SMBHBs modeled as,
\begin{align}
    h^2\Omega_{\scriptscriptstyle\textrm{GW}}(f) = \frac{2 \pi^2A_{\scriptscriptstyle\textrm{BHB}}^2}{3 H_0^2} \left(\frac{f}{\textrm{year}^{-1}}\right)^{5-\gamma_{\scriptscriptstyle\textrm{BHB}}}\textrm{year}^{-2}\,.
\end{align}
Particularly, we use {\texttt{bhb\_th\_prior}}\texttt{=True} option and choose 2D Gaussian prior for 
$\gamma_{\scriptscriptstyle\textrm{BHB}}$ and $A_{\scriptscriptstyle\textrm{BHB}}$,
derived by performing a power-law fit to the simulated SMBHB populations \cite{NANOGrav:2023hvm, Mitridate:2023oar}.

We use the PBH mass $\MPBH$, PBH domination duration $\tau_{\rm rat}$, and the slope of the blue tilted part of the scalar spectra $n_0$ as input parameters for our MCMC runs. We also obtain $\beta_f$ as a derived parameter, a function of $\MPBH$ and $\tau_{\rm rat}$. The $\MPBH$ and $\beta_f$ determine the peak wavenumber $k_f$ associated with the PBH formation peak, as discussed in Eqn. (\ref{waves}). Thus, the inflationary power spectrum is determined by this peak value $k_{\rm pk}=k_f$, the slope of the blue tilted part $n_0$, and the height of the peak $A_0$, which is determined using Eqn. (\ref{ps01}) to get the corresponding value of $\beta_f$. 

One interesting point in this context is the effect of the chosen PBH collapse criteria. The calculation of $\beta_f$ from the scalar power spectra is exponentially sensitive to this choice.
Thus, after fixing the primordial scalar spectra, one would expect a large uncertainty in the resonant GW production due to the uncertainties in the estimated $\beta_f$, as $\beta_f$ determines the duration of PBH domination and the resonant SGWB is highly sensitive to this duration. However, our analysis traces this path from the opposite direction (as shown in Fig. \ref{flc}), leading to a suppressed sensitivity towards the uncertainties in PBH formation conditions. Our input parameters are the PBH mass $M_{\rm PBH}$ and the duration of PBH-dominated era $\tau_{\rm rat}$ which determine the required PBH abundance $\beta_f$ \cite{Bhaumik:2022pil, Bhaumik:2022zdd}. Using  $\beta_f$, we determine the height of the scalar power spectra parameter $A_0$.
While the calculation of $\beta_f$ from $A_0$ is exponentially sensitive to the value of $\delta_c$, that is simply not the case for the inverse calculation (from $\beta_f$ to $A_0$). For a fixed $\beta_f$, the variance of density contrast $\sigma_\delta$ is only linearly sensitive to the changes in $\delta_c$, and thus the changes in PBH collapse condition can only lead to small changes in $A_0$ and resulting $\Omega_{\rm GW}$.

During the data analysis, we carefully consider following theoretical and observational bounds and exclude the inconsistent regions of parameter space.

\begin{itemize}
\item {\bf PBH evaporation bound}: Depending on initial PBH abundance and mass range, PBH can either dominate the universe briefly or evaporate before they can dominate. We set $\tau_{\rm rat} \gg 1 $ to avoid the parameter region where PBHs evaporate before they can dominate.
\item { \bf $\Delta N_{\rm eff}$ bound}: GWs generated before BBN act as an extra relativistic component, which both BBN and CMB observations severely constrain in terms of the effective number of neutrinos, $\Delta N_{\rm eff}$ \cite{Smith:2006nka, Caprini:2018mtu}, thereby restricting our parameter space,
\be
\Omega_{\rm GW} h^2 \rvert_{\rm peak} \lesssim 6.9 \times 10^{-6} \, .
\ee

\item {\bf Bound from CMB scales}: We limit our parameter search to the region where the scalar spectrum is unaffected at CMB scales. We ensure that at $k= 1~{\rm Mpc}^{-1}$, 
\be
A_s \left(\frac{k}{k_{\rm p}}\right)^{n_s-1} \ge A_0 \left(\frac{k}{k_{\rm pk}}\right)^{n_0-1}  \,.
\ee
\end{itemize}

Figure~\ref{fig:scan} shows the results of our scan over the model parameter space for priors described in Table \ref{tab:priors}. We find a good fit to the NANOGrav data for the PBH mass $M_{\rm PBH}$ between $6.9 \times 10^{7}$g and $1.3 \times 10^{8}$g while the abundance $\beta_f$ lies between $2.0 \times 10^{-10}$ and $3.8 \times 10^{-10}$ with the spectral index $n_0$ in a narrow range $1.48$ to $1.53$. The results for IPTA are similar, with the ranges slightly increased. The mean values of $\MPBH$, $n_0$, and $\beta_f$ obtained from the scan are specified in Table~\ref{tab:meanPoseterior} and the Bayes factor with respect to the SMBHB model in Table~\ref{tab:BF}. We find the Bayes factor larger than unity for both the NG15 and IPTA2 datasets.

In the left panel of Fig.~\ref{fig3}, we display the inflationary scalar power spectra with orange lines for the posterior mean values from our scan with NANOGrav 15 years dataset (see Table~\ref{tab:meanPoseterior}). 
The spectrum is within reach of PIXIE~\cite{2011JCAP...07..025K} and super-PIXIE~\cite{Kogut:2019vqh} which could verify our scenario in the near future~\cite{2021ExA....51.1515C}.

In the right panel of Fig.~\ref{fig3}, we show the GW spectra associated with the same inflationary power spectrum as well as projected sensitivities of
LIGO/Virgo/KAGRA (LVK)~\cite{LIGOScientific:2014pky, LIGOScientific:2016fpe, LIGOScientific:2019vic} at the end of its operation, the Einstein Telescope (ET)~\cite{Punturo:2010zz, Hild:2010id}, LISA~\cite{Bartolo:2016ami, Caprini:2019pxz, LISACosmologyWorkingGroup:2022jok}, the Nancy Roman telescope (ROMAN)~\cite{Wang:2022sxn} and two atom interferometry experiments AION~\cite{Badurina:2019hst} and AEDGE~\cite{AEDGE:2019nxb}.
The grey violins show the recent NANOGrav data~\cite{NANOGrav:2023gor, NANOGrav:2023hde} with some of the widest bins that do not contribute very significantly to the fit omitted for clarity. 
We see that the solid orange line associated with the resonant peak produced by our scalar perturbation spectrum fits the PTA data very well and would be within reach of astrometry experiments such as the Nancy Roman Telescope.
The dashed lines represent second-order SGWB associated with the production of PBHs. As we see, the third-generation terrestrial laser interferometers such as the Einstein Telescope (ET)~\cite{Punturo:2010zz, Hild:2010id} or Cosmic Explorer~\cite{Reitze:2019iox}  will be able to probe our scenario as expected~\cite{Franciolini:2023osw}. We also verified the same holds for all the points in our scan; they are all within reach of ET but not of any of the other indicated experiments. 

The detection of this second SGWB peak or the PBH formation peak can play a very crucial role in identifying the curvature power spectrum in the broader $k$ range. For example, if we take an optimally chosen power-law for $k<k_m$ with a sharp peak at $k \sim k_f$, we can obtain identical results for the resonant SGWB peak satisfying PTA data. 
However, this degeneracy can be easily broken with the PBH formation peak, as both the shape and the amplitude of the PBH formation peak would change significantly if we consider a different form of the scalar 
power spectra near $k \sim k_f$.

{\bf Discussion and future prospects:}.
The recent detection of SGWB in PTAs has triggered an intensive search for possible cosmological sources that can explain the signal. In this context, the second-order SGWB generated during the formation of comparatively massive PBHs has been considered in the literature already \cite{NANOGrav:2023hvm, Inomata:2023zup, Franciolini:2023pbf}. 
In this paper, we focus on the resonant SGWB associated with the domination of ultra-low mass PBHs, in the early universe. 
In particular, we assume a broadly peaked inflationary scalar power spectrum leads to the formation of the PBHs that dominate the expansion briefly before evaporating. The non-trivial evolution of scalar perturbations during that period contributes to the resonant amplification of SGWB upon return to RD.  We find that this background can adequately explain the observed NANOGrav signal.
We compare our ultra-low mass PBH scenario with the SMBHB merger model and find Bayesian evidence in favor of our model for NG15 and IPTA2 datasets.

While the resonant SGWB in the scenario we consider is consistent with the PTA observations, there are also two other methods to verify this possibility.
Firstly, the SGWB associated with the formation of these ultra-low mass PBHs would be within the reach of third-generation terrestrial laser interferometers such as ET, as shown in the right side of Fig.~\ref{fig3}. Secondly, the CMB spectral distortion probes like PIXIE~\cite{2011JCAP...07..025K, 2012ApJ...758...76C}  would be able to probe the broadly-peaked inflationary scalar spectra scenario in near future~\cite{2021ExA....51.1515C}, as we show in the left panel of Fig.~\ref{fig3}.

\begin{acknowledgments}
\vspace{5pt}\noindent\emph{Acknowledgments:}
The authors thank Theodoros Papanikolaou for raising interesting points and useful discussion. This work was supported by the Polish National Agency for Academic Exchange within Polish Returns Programme under agreement PPN/PPO/2020/1/00013/U/00001 and the Polish National Science Center grant 2018/31/D/ST2/02048. RKJ acknowledges financial support from the new faculty seed start-up grant of the Indian Institute of Science, Bengaluru; Science and Engineering  Research Board, Department of Science and Technology, Govt. of India, through the Core Research Grant~CRG/2018/002200, the MATRICS grant~MTR/2022/000821 and the Infosys Foundation, Bengaluru, India, through the Infosys Young Investigator award.
\end{acknowledgments}

\bibliographystyle{apsrev4-1}
\bibliography{apssamp}

\end{document}